# An Analytical Method for Calculating the Satellite Bow Shock/Magnetopause Interception Positions and Times

Atanas Marinov Atanassov

*Solar Terrestrial Influences Institute, Bulgarian Academy of Sciences,
Stara Zagora Department, P.O. Box 73, 6000 Stara Zagora, Bulgaria*

**Summary**
This paper contains a presentation of analytical solution of the problem of calculating the places and moments of intersection of satellite trajectories with elements of the Earth's magnetosphere (bow shock and magnetopause). The satellite motion is presented in a Kepler's approximation. Magnetopause and bow shock are described by second-order surfaces- elliptic paraboloides. These surfaces are employed as situational conditions for determining the points of intersection they have (if any) with the satellite trajectory. The situational condition is herein transformed into the plane of Kepler's orbit, thereafter it is reduced to a second-order plane curve- quadric (ellipse or parabola). The solution of this system, containing the equation of this curve and Kepler's ellipse equation, allows determining the places where orbits intersect with the magnetopause or the bow shock. The solution of this system is suggested to be given by reducing the system to a fourth-order equation.

## 1. Introduction

The bow shock (BS) and the magnetopause (MP) are the basic elements of the Earth magnetosphere [1]. BS is the outer boundary thin skin of the magnetosphere where the solar wind decelerates and its parameters (velocity, pressure) are strongly changed by interaction with the Earth magnetic field. MP is the actual boundary between the shocked solar wind and the magnetospheric plasma where the physical processes at this boundary control the entry of plasma, momentum, energy and the redistribution of geomagnetic flux.

When satellite experiments are designed for investigating the Earth's magnetosphere, it is important to detect the places and moments of intersection of some elements such as BS and MP. This is the basis on which the satellite orbits are optimized [2]. Besides, during the planning stage of the experiment, the operational modes of any devices (detectors of particles or fields) that conduct measurements in the respective time intervals are determined.

The position of the Earth's bow shock and magnetopause are approximated well enough by employing second-order surfaces (quadric):

$$a_1 x^2 + a_2 y^2 + a_3 z^2 + a_4 xy + a_5 xz + a_6 yz + a_7 x + a_8 y + a_9 z + a_{10} = 0 \qquad (1)$$

The equations of these surfaces have constant coefficients in the geocentric solar-ecliptic coordinate system (described below) [3, 4]:
bow shock-
$$0.03x^2 + y^2 + 1.01z^2 + 0.16xy - 0.02xz - 0.09yz + 43.48x - 4.00y - 0.88z - 615 = 0$$

magnetopause-
$$0.37x^2 + y^2 + 0.98z^2 + 0.09xy - 0.23xz - 0.04yz + 17.94x - 1.20y - 0.48z - 230 = 0$$

The analyses presented in [2] are based on determining the satellite trajectories by a numerical integration of the equation of motion and transformation of the coordinates within a coordinate system, where the coefficients of (1) are known.

In [5] is presented a relatively general approach to a situational analysis, which allows covering various cases where the situational condition can be presented by a geometrical model (various parameterizable surfaces). The section of the Kepler's plane with a surface is a planar curve. Determining of this curve is equivalent to a transformation of the situational condition in the plane of Kepler's orbit. The points of intersection of this curve with Kepler's orbit determine the satellite location and time within the area determined by the particular situational condition. The employment of this approach does not require to discretize the orbit neither to examine the situational condition for various time moments. The requirement related to the applicability of this approach is to have constant situational conditions at the time intervals under examination. So as to be able to use the Kepler's approximation for presenting the satellite orbital motion, the modifications of the satellite orbital elements should be minor enough too.

This paper presents the transformation of the situational condition by the means of an equation (1) to a new equation, describing a curve in the Keplerian plane. The last equation, together with the Kepler's equation, forms a second-degree system with two unknowns. This system is transformed into one fourth-degree algebraic equation with one unknown. The solution of this equation allows determining the points of intersection of the satellite orbit and the BS or MP. After the points of intersection are identified by the means of Kepler's equation, then the moments when this is to happen are to be identified.

**2. Coordinate Systems**

Some of the coordinate systems to be used in this paper include:
- **Geocentric Equatorial Inertial System (GEI)** [6]; the origin of this coordinate system is the centre of the Earth, the X-axis points towards the point of the autumnal equinox, the Z-axis is parallel to the Earth rotation axis and the Y-axis complements to a right handed orthogonal set.
- **Geocentric Solar Ecliptic System (GSE)** [7]; this system has the same origin as the **GEI**, the X-axis points towards the Sun, the Z-axis is perpendicular to the ecliptic plane. The angle between the equatorial and ecliptic planes is $\varepsilon = 23° 27'$. The angular velocity of the motion of the Sun along the ecliptic is $\lambda' \cong 1°$/day. Since the X-axis changes its direction, GSE is mobile.
- **Orbital System (OS)** [4]; the origin of this system is one of the focuses of Kepler's orbit, which is identical with the **GEI** center, the X-axis points towards the pericenter, the Y-axis is perpendicular to X-axis and points towards the direction of the real anomaly increase in the first quadrant, while the Z-axis is perpendicular to the orbital plane and supplements the first two axes to a right handed orthogonal set. The orbital plane orientation can be determined based on the length of the ascending node $\Omega$ and the inclination **i** towards the equatorial plane; the argument of perigee $\omega$ describes the orientation of the orbit in Kepler's plane. Any changes in the orbital elements change the orientation of **OS**.

This paper uses **GSE** to define the position of the Sun, but also the positions of objects that are thereto connected, namely BS and MP. **OS** is used to describe the orbital motion. **GEI** is used as a link between the other two coordinate systems.

### 3. Transformation of the Situational Condition Towards the Plain of Kepler's Orbit.

The transformation of the orbital coordinates into solar-ecliptic ones, can be expressed as follows:

$$\left\| \begin{array}{c} x \\ y \\ z \end{array} \right\| = \alpha \cdot \beta \left\| \begin{array}{c} \xi \\ \eta \\ \zeta \end{array} \right\|, \qquad (2)$$

where $\alpha$ and $\beta$ are transformation matrix from **GEI** into **GSE** and from **OS** into **GEI** respectively.

By multiplying the transformation matrices $\alpha_{SEGE}$ и $\beta_{GEOS}$, we obtain:

$$\begin{vmatrix} x = \gamma_{11}\xi + \gamma_{12}\eta + \gamma_{13}\zeta \\ y = \gamma_{21}\xi + \gamma_{22}\eta + \gamma_{23}\zeta \\ z = \gamma_{31}\xi + \gamma_{32}\eta + \gamma_{33}\zeta \end{vmatrix} \qquad (3)$$

In (3) $\gamma_{ij}$ are elements of the transformation matrix between **OS** and **GEI**:

$\gamma_{11}=$ $cos\lambda$ ($cos\omega$ $cos\Omega$ – $sin\omega$ $sin\Omega$ $cos$i) + $sin\lambda$ $cos\varepsilon$ ($cos\omega$ $sin\Omega$ + $sin\omega$ $cos\Omega$ $cos$i) +
+ $sin\lambda$ $sin\varepsilon$ $sin\omega$ $sin$i

$\gamma_{12}=$ –$cos\lambda$ ($sin\omega$ $cos\Omega$ + $cos\omega$ $sin\Omega$ $cos$i) – $sin\lambda$ $cos\varepsilon$ ($sin\omega$ $sin\Omega$ – $cos\omega$ $cos\Omega$ $cos$i) +
+ $sin\lambda$ $sin\varepsilon$ $cos\omega$ $sin$i

$\gamma_{13}=$ $cos\lambda$ $sin\Omega$ $sin$i – $sin\lambda$ $sin\varepsilon$ $cos\Omega$ $sin$i + $sin\lambda$ $sin\varepsilon$ $cos$i

$\gamma_{21}=$ –$sin\lambda$ ($cos\omega$ $cos\Omega$ – $sin\omega$ $sin\Omega$ $cos$i) + $cos\lambda$ $cos\varepsilon$ ($cos\omega$ $sin\Omega$ + $sin\omega$ $cos\Omega$ $cos$i) –
+ $cos\lambda$ $sin\varepsilon$ $sin\omega$ $sin$i

$\gamma_{22}=$ $sin\lambda$ ($sin\omega$ $cos\Omega$ + $cos\omega$ $sin\Omega$ $cos$i) – $cos\lambda$ $cos\varepsilon$ ($sin\omega$ $sin\Omega$ – $cos\omega$ $cos\Omega$ $cos$i) –
+ $cos\lambda$ $sin\varepsilon$ $cos\omega$ $sin$i

$\gamma_{23}=$ –$sin\lambda$ $sin\Omega$ $sin$i – $cos\lambda$ $cos\varepsilon$ $cos\Omega$ $sin$i + $cos\lambda$ $sin\varepsilon$ $cos$i

$\gamma_{31}=$ –$sin\varepsilon$ ($cos\omega$ $sin\Omega$ + $sin\omega$ $cos\Omega$ $cos$i) + $cos\varepsilon$ $sin\omega$ $sin$i

$\gamma_{32}=$ $sin\varepsilon$ ($sin\omega$ $sin\Omega$ – $cos\omega$ $cos\Omega$ $cos$i) + $cos\varepsilon$ $cos\omega$ $sin$i

$\gamma_{33}=$ $sin\varepsilon$ $cos\Omega$ $sin$i + $cos\varepsilon$ $cos$i

By replacing x, y, and z from (3) in (1) and after the relevant transformations, with respect to the equation of second-order surfaces transformed into the orbital coordinate system, we produce:

$$B_1\xi^2 + B_2\eta^2 + B_3\zeta^2 + B_4\xi\eta + B_5\xi\zeta + B_6\eta\zeta + B_7\xi + B_8\eta + B_9\zeta + a_{10} = 0, \qquad (4)$$

where the coefficients $B_i$ assume the following expression:

$$B_1 = a_1\gamma_{11}^2 + a_2\gamma_{21}^2 + a_3\gamma_{31}^2 + a_4\gamma_{11}\gamma_{21} + a_5\gamma_{11}\gamma_{31} + a_6\gamma_{21}\gamma_{31}$$

$$B_2 = a_1\gamma_{12}^2 + a_2\gamma_{22}^2 + a_3\gamma_{32}^2 + a_4\gamma_{12}\gamma_{22} + a_5\gamma_{12}\gamma_{32} + a_6\gamma_{22}\gamma_{32}$$

$$B_3 = a_1\gamma_{13}^2 + a_2\gamma_{23}^2 + a_3\gamma_{33}^2 + a_4\gamma_{13}\gamma_{23} + a_5\gamma_{13}\gamma_{33} + a_6\gamma_{23}\gamma_{33}$$

$$B_4 = 2a_1\gamma_{11}\gamma_{12} + 2a_2\gamma_{21}\gamma_{22} + 2a_3\gamma_{21}\gamma_{32} + a_4(\gamma_{11}\gamma_{22} + \gamma_{12}\gamma_{21}) + a_5(\gamma_{11}\gamma_{32} + \gamma_{12}\gamma_{31}) \\ + a_6(\gamma_{21}\gamma_{32} + \gamma_{22}\gamma_{31})$$

$$B_5 = 2a_1\gamma_{11}\gamma_{13} + 2a_2\gamma_{21}\gamma_{23} + 2a_3\gamma_{31}\gamma_{33} + a_4(\gamma_{11}\gamma_{23} + \gamma_{13}\gamma_{21}) + a_5(\gamma_{11}\gamma_{13} + \gamma_{13}\gamma_{31}) \\ + a_6(\gamma_{21}\gamma_{33} + \gamma_{23}\gamma_{31})$$

$$B_6 = 2a_1\gamma_{12}\gamma_{13} + 2a_2\gamma_{22}\gamma_{23} + 2a_3\gamma_{32}\gamma_{33} + a_4(\gamma_{12}\gamma_{23} + \gamma_{13}\gamma_{22}) + a_5(\gamma_{12}\gamma_{33} + \gamma_{13}\gamma_{32}) \\ + a_6(\gamma_{22}\gamma_{33} + \gamma_{23}\gamma_{32})$$

$$B_7 = a_7\gamma_{11} + a_8\gamma_{21} + a_9\gamma_{31}$$

$$B_8 = a_7\gamma_{12} + a_8\gamma_{22} + a_9\gamma_{32}$$

$$B_9 = a_7\gamma_{13} + a_8\gamma_{23} + a_9\gamma_{33}$$

The equation of the section of Kepler's plane with the second-order surface in question, transformed towards the **OS,** is produced by substituting $\zeta=0$, or:

$$B_1\xi^2 + B_2\eta^2 + B_4\xi\eta + B_7\xi + B_8\eta + a_{10} = 0 \qquad (5)$$

A second-order curve is presented in (5) which defines the situational condition into the plane of Kepler's orbit - **situational quadric (SQ)**, as shown in Fig.1. Depending on the coefficients $B_i$, this curve may assume different meanings in terms of geometry. The second invariant of this curve has the following expression [8]:

$$I_2 = B_1B_4 - B_2B_2.$$

Property of this function is to remain constant when the coordinate system is transformed. The value of this invariant defines what the type for the curve to take is:

$I_2 > 0$ – ellipse,

$I_2 < 0$ - hyperbola,

$I_2 = 0$ – parabola.

Given that (4) is a parabolic surface equation, then $I_2 \geq 0$.

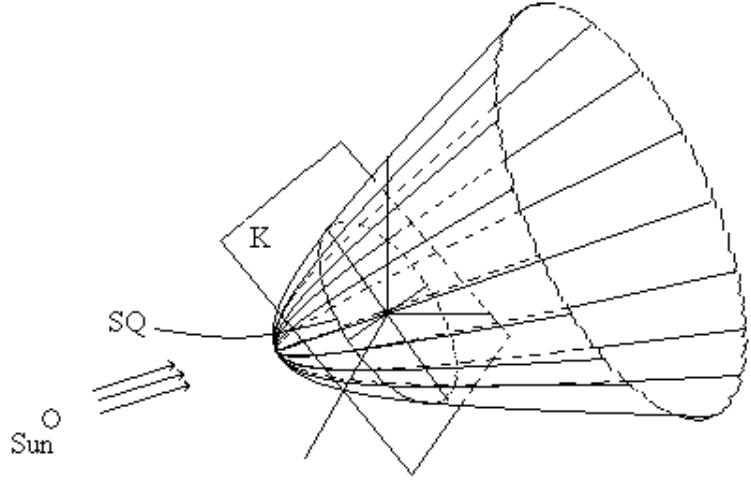

Fig. 1 K- Kepler's plane, situational quadric SQ- section between parabolic surface and parabolic surface.

**4. Determination of the Points of Intersection of Kepler's Ellipse with the Situation Condition Curve**

On the other hand, Kepler's ellipse equation in the orbital coordinate system appears as follows [6]:

$$\frac{(\xi+c)^2}{a^2}+\frac{\eta^2}{a^2(1-e^2)}=1, \quad c = a.e \tag{6}$$

where **a** is the semi-major axis and **e** is the eccentricity.

Solving the equations (5) and (6) together, helps us determine the points of intersection of Kepler's ellipse with a surface which approximates SB and MP. First of all, let's rewrite this equations system as follows:

$$\begin{Vmatrix} \xi^2 + L_2\eta^2 + L_3\xi\eta + L_4\xi + L_5\eta + L_6 = 0 \\ \xi^2 + 2c\xi + s\eta_2 + c^2 = r \end{Vmatrix} \tag{7}$$

$$L_i = \frac{B_i}{B_1}, 1 < i \le 5; L_6 = \frac{a_{11}}{B_1}; r = a^2; s = \frac{1}{1-e^2}$$

Extracting the second equation from the first one and after some transformation, we produce:

$$\xi = \frac{R - P\eta^2 - L_5\eta}{L_3\eta + Q}, \tag{8}$$

were $R = c^2 - r - L_6$, $P = L_2 - s$, $Q = L_4 - 2c$. Replacing with the expression obtained for $\xi$ in the second equation of (7) and after the relative transformations, we can express this as:

$$C_1\eta^4 + C_2\eta^3 + C_3\eta^2 + C_4\eta + C_5 = 0, \tag{9}$$

where:

$$C_1 = P^2 + sL_3^2$$

$$C_2 = 2PL_5 + 2sL_3Q - 2cL_3P$$

$$C_3 = L_5^2 - 2RP - 2cPQ - 2cL_3L_5 + sQ^2 + 2QL_3\varpi$$

$$C_4 = -2RL_5 + 2cRL_3 - 2cQL_5 + 2L_3Q\varpi$$

$$C_5 = 2cRQ + 2c + Q^2\varpi; \qquad \varpi = c^2 - r$$

The equation (9) is algebraic fourth-order equitation. There are two methods that can be applied to present the solution of this equation in analytical way. The first one is the Descartes-Euler's method and the second one is the Ferrari's. Different numerical methods can be also applied. These methods help finding the roots $\bar{\eta}_1$, $\bar{\eta}_2$, $\bar{\eta}_3$ and $\bar{\eta}_4$ of the equation (9), then we come back to the system of equations. We will not focus on analyzing the roots of pointed quartic neither on comparing the different solving methods.

The number of the real roots determines the number of the points of intersection between Kepler's ellipse and the second-order curve (ellipse or parabola), as illustrated in Fig. 2. By replacing with the calculated values for $\bar{\eta}$ in one of the equations (7) (let's take the simplest one), we find the four corresponding values of $\bar{\xi}_{n;n=\overline{1,4}}$. The four couples of values thus calculated, namely $(\bar{\xi}_1, \bar{\eta}_1)$, $(\bar{\xi}_2, \bar{\eta}_2)$, $(\bar{\xi}_3, \bar{\eta}_3)$ and $(\bar{\xi}_4, \bar{\eta}_4)$, present the system roots. The full analysis gives two or four points of interception.

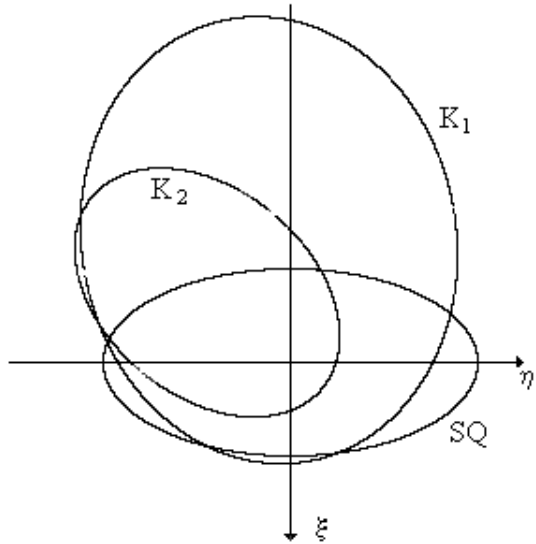

Fig. 2. Two Kepler's ellipses $K_1$ and $K_2$ intercept situational quadric SQ in forth and two points and respectively.

## 5. Determination of the Moments When a Satellite Intersects with the Bow Shock/Magnetopause.

The following relation between the orbital coordinates and the eccentric anomaly E [6] exists:

$$\begin{cases} \xi = a(\cos E - e) \\ \eta = a\sqrt{1-e^2} \cdot \sin E \end{cases} \quad (10)$$

On the other hand, on the base of Kepler's equation, the following expression can be produced:

$$t = t_0 + (E - e \cdot \sin E)/\lambda \quad (11)$$

By determining the eccentric anomaly E from (10) and substituting in (11) accordingly, we determine the moments when the satellite intersects with the MP or BS.

## 6. Conclusion

This paper presents an approach for determining the places and moments at which the satellite orbits intersect with MP and SB. It is based on the requirement that the situational condition and the orbit elements are steady. In reality, this is feasible at determined time intervals $\Delta t$ when the coefficients of equation (1) are constant and the change of BS and MP orientation caused by the motion of the Earth around the Sun is negligible. The variations of the satellite orbital elements at the time interval $\Delta t$ should be negligible enough.

It is necessary to carry out an error analysis related to the determination of the places and moments of the intersection of satellite orbits with MP and BS based on the approach herein presented. A study related with the potential of the various methods for solving system (7) and equation (9) will be carried out in the future.